\input{epsf}
\documentstyle[twoside]{article}
\catcode`\@=11
\long\def\@makefntext#1{ %\parindent 1em
\protect\noindent \hbox to 3.2pt {\hskip-.9pt
$^{{\eightrm\@thefnmark}}$\hfil}#1\hfill} %can be used

 \def\@makefnmark{\hbox to 0pt{$^{\@thefnmark}$\hss}}  %original

\def\ps@myheadings{\let\@mkboth\@gobbletwo
\def\@oddhead{\hbox{} %\sl
\rightmark\hfil\eightrm\thepage}
\def\@oddfoot{}\def\@evenhead{\eightrm\thepage\hfil %\sl
\leftmark\hbox{}}\def\@evenfoot{}
\def\sectionmark##1{}\def\subsectionmark##1{}}
\oddsidemargin=\evensidemargin
\newcounter{sectionc}\newcounter{subsectionc}\newcounter{subsubsectionc}
\renewcommand{\section}[1] {\vspace{8pt}\addtocounter{sectionc}{1}
\setcounter{subsectionc}{0}\setcounter{subsubsectionc}{0}\noindent
        {\elevenbf\thesectionc. #1}\par\vspace{4pt}}
\renewcommand{\subsection}[1] {\vspace{6pt}\addtocounter{subsectionc}{1}
        \setcounter{subsubsectionc}{0}\noindent
        {\bf\thesectionc.\thesubsectionc. {\kern1pt \bf #1}}\par\vspace{5pt}}
\renewcommand{\subsubsection}[1]
{\vspace{12pt}\addtocounter{subsubsectionc}{1}
        \noindent{\tenrm\thesectionc.\thesubsectionc.\thesubsubsectionc.
        {\kern1pt \tenit #1}}\par\vspace{5pt}}

\newcommand{\bibit}{\nineit}
\newcommand{\bibbf}{\ninebf}
\renewenvironment{thebibliography}[1]                   %ALL CHANGES DD 13/3/92
        {\ninerm
         \baselineskip=11pt                             %changed by cheng
         \begin{list}{\arabic{enumi}.}
        {\usecounter{enumi}\setlength{\parsep}{0pt}
         \setlength{\leftmargin 17pt}{\rightmargin 0pt} %changed by cheng
         \setlength{\itemsep}{0pt} \settowidth          %changed by cheng
        {\labelwidth}{#1.}\sloppy}}{\end{list}}
\def\@citex[#1]#2{\if@filesw\immediate\write\@auxout    %IJMPA, IJMPB ONLY
        {\string\citation{#2}}\fi                       %TO DELETE PERCENTAGE
\def\@citea{}\@cite{\@for\@citeb:=#2\do                 %KEY WHEN USING
        {\@citea\def\@citea{,}\@ifundefined             %DD 13/3/92
        {b@\@citeb}{{\bf ?}\@warning
        {Citation `\@citeb' on page \thepage \space undefined}}
        {\csname b@\@citeb\endcsname}}}{#1}}

\newif\if@cghi
\def\cite{\@cghitrue\@ifnextchar [{\@tempswatrue
        \@citex}{\@tempswafalse\@citex[]}}
\def\citelow{\@cghifalse\@ifnextchar [{\@tempswatrue
        \@citex}{\@tempswafalse\@citex[]}}
\def\@cite#1#2{{$\null^{#1}$\if@tempswa\typeout
        {IJCGA warning: optional citation argument
        ignored: `#2'} \fi}}

\font\tenit=cmti10
\font\tenrm=cmr10

\font\elevenbf=cmbx10 scaled\magstep 1
 1
 1
\font\ninebf=cmbx9
\font\ninerm=cmr9
\font\nineit=cmti9

\font\eightrm=cmr8

\newcommand{\be}{\begin{equation}}
\newcommand{\ee}{\end{equation}}
\newcommand{\bea}{\begin{eqnarray}}
\newcommand{\eea}{\end{eqnarray}}
\font\oe=cmoer10

\def\folio{\ifnum\pageno>0\hbox{\number\pageno}\fi}
\def\text[#1]{\hbox{\rm #1}}
\def \etal{{\em et. al\/}}

\def\R(#1){\cite{#1}}
\def\E(#1){Eq.\/(\ref{eq:#1})}
\def\Fig(#1){Fig.\/(\ref{f:#1})}
\def\SS(#1){\SS\thinspace#1}

\def\sf(#1,#2){\textstyle{#1\over#2}\displaystyle} % small fraction

\def\w{\widetilde}
\def\t{\tilde}

\def\b#1{\bar{#1}}

\mathchardef\eth="0164
\mathchardef\thn="0175
\def\edth{\hbox{$\textfont1=\oe {\eth}$}}

\def\AT{A_{\cal T}}
\renewcommand{\H}{\cal H}
\def\Hm{\cal H^{-}}
\def\L{\Lambda}

\def\N{N}

\def\P0{{P_0}}

\def\Scri#1{{\cal J}^{#1}}
\def\Sinf{{{\cal S}_\infty}}
\def\Sur{\S_{u,r}}
\def\S{{\cal S}}
\newcommand{\Tm}{\cal T^{-}}
\newcommand{\Tu}{{\cal T}_u}
\newcommand{\Tmu}{{{\cal T}_{u}}^{-}}

\def\V{{\Re}}
\def\Z{Z}
\def\bl0{\Delta_{0}}
\def\ddu {\frac{d}{du}}
\def\e0{\edth_0}

\def\ints2{\int_{S^2}}
\def\mb{{\bar m}}
\def\mb{{\bar m}}
\def\q2{\gamma_0 Q \overline Q}
\def\scri+{{\cal J}^+}
\def\vol#1{\omega_{#1}}
\def\zb{{\bar\zeta}}
\def\z{{\zeta}}

\def\RT{Robinson-Trautman }
\def\RTEM{Robinson-Trautman Einstein-Maxwell }
\def\af{asymptotically flat }

\def\JournalRef.#1.#2.#3.#4.{(19#4) {\bibit\/#1\ }{\bibbf #2}, #3.}
\def\ANYAS.#1.#2.#3.{\JournalRef.{Ann. NY Acad. Sci.}.#1.#2.#3.}
\def\CMP.#1.#2.#3.{\JournalRef.{Commun. Math. Phys. }.#1.#2.#3.}
\def\CQG.#1.#2.#3.{\JournalRef.{Class. Quantum. Grav.}.#1.#2.#3.}
\def\GRG.#1.#2.#3.{\JournalRef.{Gen. Rel. Grav.}.#1.#2.#3.}
\def\IJMP.#1.#2.#3.{\JournalRef.{Int. J. Mod. Phys.}.#1.#2.#3.}
\def\JMP.#1.#2.#3.{\JournalRef.{J. Math. Phys.}.#1.#2.#3.}
\def\NC.#1.#2.#3.{\JournalRef.{Nuovo Cimento}.#1.#2.#3.}
\def\PLet.#1.#2.#3.{\JournalRef.{Phys. Lett.}.#1.#2.#3.}
\def\PR.#1.#2.#3.{\JournalRef.{ Phys. Rev.}.#1.#2.#3.}
\def\PRD.#1.#2.#3.{\JournalRef.{Phys. Rev. D}.#1.#2.#3.}
\def\PRL.#1.#2.#3.{\JournalRef.{Phys. Rev. Lett.}.#1.#2.#3.}
\def\PRSL.#1.#2.#3{\JournalRef.{Proc. Roy. Soc. Lond.}.#1.#2.#3.}
\def\PRep.#1.#2.#3.{\JournalRef.{Phys. Rep.}.#1.#2.#3.}
\def\RMP.#1.#2.#3.{\JournalRef.{Rev.Mod. Phys.}.#1.#2.#3.}
\def\A.#1.#2.#3.{\JournalRef.{   }.#1.#2.#3.}

\def\BookRef.#1.#2.{(19#2, #1)}
\def\CUP.#1.{\BookRef.{Cambridge University Press, Cambridge}.#1.}
\def\DOV.#1.{\BookRef.{Dover, New York}.#1.}
\def\KAP.#1.{\BookRef.{Kluwer Academic Publishers, Dordrecht}.#1.}
\def\MGH.#1.{\BookRef.{McGraw-Hill, New York}.#1.}
\def\OUP.#1.{\BookRef.{Oxford University Press, Oxford}.#1.}
\def\SIAM.#1.{\BookRef.{SIAM, Philadelphia}.#1.}
\def\SV.#1.{\BookRef.{Springer-Verlag, Berlin}.#1.}
\def\WS.#1.{\BookRef.{World Scientific, Singapore}.#1.}

\begin{document}
%\normalsize\textlineskip
%{\thispagestyle{empty}
\setcounter{page}{1}
\textheight=7.8truein
\setcounter{footnote}{0}

\title{Apparent Horizons in Vacuum Robinson-Trautman Spacetimes
\thanks{Talk given at the inaugural Australian General Relativity Workshop,
A.N.U, Canberra, 26th September 1994.}}
\author{E.W.M.Chow
	 and A.W.-C.Lun \\
		Department of Mathematics, Monash University
		\thanks{E-mail: ewmc@cosmo.maths.monash.edu.au,
		lun@vaxc.cc.monash.edu.au.
		Postal address: Wellington Rd, Clayton 3168 Australia. }}
\date{30th March 1995
		\thanks{This preprint is available from the above address as
		Applied Mathematics Preprint 8/95 or from the xxx.lanl.gov e-print
		archive as gr-qc/9503065.}}
\maketitle

\begin{abstract}
Vacuum asymptotically flat Robinson-Trautman spacetimes are a well known class
of spacetimes exhibiting outgoing gravitational radiation.  In this paper we
describe a method of locating the past apparent horizon in these spacetimes,
and discuss the properties of the horizon.  We show that the past apparent
horizon is non-timelike, and that its surface area is a  decreasing function
of the retarded time.  A numerical simulation of the apparent horizon is also
discussed.
\end{abstract}
\section{Introduction}
Vacuum asymptotically flat Robinson-Trautman spacetimes have been the subject
of much study since their discovery over thirty years ago.\R(RT60) They
possess some very nice features which make them amenable to analysis.  They
are in some sense the simplest asymptotically flat solutions which exhibit
gravitational radiation, albeit of a fairly specialised nature.  The feature
of the spacetime which simplifies analysis is that the full spacetime can be
built up from the solution of a fourth order parabolic equation on a 2+1
dimensional manifold.  This is a consequence of the fact that the coordinate
system uses a retarded time coordinate, so that in effect initial data is
prescribed on a null hypersurface.

Early studies of the \RT equation related to behaviour of solutions of the
linearised equation.\R(FN67,Van87)  Luk\'acs {\it et.al.}\R(LPPS) studied the
behaviour of solutions of the full nonlinear equation, using the concept of
Lyapunov stability to establish that global solutions, if they existed, would
converge to the static Schwarzschild equilibrium.  A number of authors
subsequently focused on the existence of such solutions.  Schmidt\R(Sch88)
showed local existence of solutions for sufficiently differentiable but
otherwise arbitrary initial data.  Rendall\R(Ren88)  showed global existence
for sufficiently small initial data, antipodally symmetric on the sphere, and
his proof was extended by Singleton\R(Sin90a)  to remove the requirement of
antipodal symmetry.  Finally Chru\'sciel\R(Chr91) proved semi-global existence
of solutions for arbitrary smooth initial data.  Numerical studies by
Perj\'es\R(Per89) and Singleton\R(Sin90b) demonstrate vividly the evolution of
an initially perturbed spacetime to the steady state.  It is interesting to
note that the evolution equations exhibit exponential divergence from
arbitrary initial data in the ``backwards'' time direction, and that no
solutions exist in this direction.\R(Chr91)  It should be emphasised these
results, and the subsequent discussion, only apply to asymptotically flat \RT
spacetimes with regular topological $S^2$ surfaces: Perj\'es and
Hoenselaers\R(HoePer) have shown that there exists a class of \RT spacetimes
with cusp singularities which evolve to static C-metrics.

Thus the evolution of the \RT spacetimes would seem to be thoroughly
understood.  We would expect then that the behaviour of physical features of
the spacetimes should not be too difficult to pin down.   Near $\Scri+$, the
spacetimes, because of their  algebraically degenerate structure, could
reasonably be interpreted as   describing purely outgoing radiation around a
black hole source: e.g. the decaying tail  of the radiation after the black
hole has formed  (Note that there is  no ingoing radiation, which means there
can be no backscattering or similar  self-interaction of the radiation
field).  However, in this paper, we will be studying the structure of past
apparent horizons in vacuum \RT spacetimes; thus we will focus our attention
on the behaviour of the ``white hole'' region of the spacetimes.  Since these
spacetimes evolve to the Schwarzschild geometry, the future apparent horizons
coincide with the future event horizons at $u=\infty$ ($r=2m$ in Schwarschild
coordinates).

Our study of the apperent horizon structure of these spacetimes was initially
motivated by an investigation of the stability of the related class of
electrovac spacetimes, the \RTEM spacetimes, discussed elsewhere.\R(LC94)
The \af vacuum \RT spacetimes turn out to have a particularly well  behaved
apparent horizon which illustrates several general  theorems  concerning
apparent horizons.  There are in fact very few well  understood exact
solutions which can be used to illustrate such theorems --  there being few
examples of non-static black hole type solutions.   The \af\RT spacetimes, in
some sense the simplest asymptotically flat spacetimes  admitting
gravitational radiation, should be a useful example for the study of
radiating black hole spacetimes in general and many of the general theorems
relating to non-stationary spacetimes.  The basic equations and features of
the spacetimes are outlined in Section 2.  Section 3 contains the results
concerning the past apparent horizon and  its properties.   Some numerical
demonstration of these results is presented in Section 4.

\vfill\pagebreak
\section {\RT vacuum spacetimes}
\subsection{Basic equations and notation}
The line element for the Robinson-Trautman class of spacetimes is given by:
      \begin{eqnarray}%%
	     ds^2 = 2Hdu^2 + 2dudr - {2r^2d\zeta d\bar\zeta \over P^2 }
	  \label{eq:metric}  \end{eqnarray}%%
where $H= -r(\ln P),_{u} + \frac{1}{2}K - \frac{m}{r}$, and $K=\Delta(lnP)$.
The vacuum \RT spacetimes are
foliated by a two parameter family of 2-surfaces $\S_{u,r}$ which, for the
asymptotically flat case, have spherical topology.
The operator $\Delta = 2P^2\partial_{\z\zb}$ is the Laplacian on these
two-surfaces.  For the vacuum spacetimes, $m=m(u)$ and we can use a coordinate
transformation on $u$ to make $m$ constant.   We are interested in the $m>0$
case, as this gives rise to a spacetime with positive Bondi mass.    The
spacetime is then determined by the evolution of $P = P(u,\z,\zb)$ on a
background two-sphere.  The evolution equation is often written as
   \begin{eqnarray}%%
            (\ln P),_u = -{1\over12m}\Delta K \label{eq:rteqn}
      \end{eqnarray}%%

This equation, the \RT equation, is also known in the
literature as the two-dimensional Calabi equation.\R(Tod89)

The steady state,
corresponding to the Schwarzschild solution, is given by
$P=P_0 = \sf(1,{\sqrt{2}})(a + b\z + \b b\zb + c\z\zb)$, where $ac-b\b{b}=1$.
Note that the equilibrium
value of $P$ is not unique but includes a freedom corresponding to conformal
motions on the sphere.  This freedom was a contributing factor to the
difficulty of proving existence of solutions of
the \RT equation.\R(Chr91)  The condition $ac-b\b b=1$ normalises the Gaussian
curvature of $S^2$ to $K=1$.

The function $P$ determines the induced 2-metric on each $\S_{u,r}$, which  is
given by
      $$ g_\S = {2r^2 d\z d\zb \over P^2} = e^{2\L}g_0 \label{eq:S2metric}$$
where $g_0$ is the metric of $S^2$, and $e^{-\L} = {P\over P_0}$.
It is possible to ``factor out'' the background $S^2$ geometry and write the
\RT equation as
      \begin{eqnarray}%%
            e^{2\L} \L,_u ={1\over 12m}\bl0{K} \label{eq:rteqn2}
      \end{eqnarray}%%
where $\bl0$ is the Laplacian on $S^2$.  We thus solve the evolution equation
on a background sphere, say at $r=1$.

\vfill\pagebreak
\subsection{Conserved quantities and Lyapunov functionals}
Several conserved integrals are implied by the evolution equations.
First, we have the conservation of surface area of $\Sur$,
and the conservation of the ``irreducible mass'', as a consequence of this:

      $$ A_\S = \int_{\Sinf}{\vol1} = \int_{S^2}{e^{2\L}\vol0} = 4\pi$$
	  $$ M_I = {1\over 4\pi} \int_{\Sinf}{m\vol1} = {m\over 4\pi}
	  \int_{S^2}{e^{2\L}\vol0}= m.\label{eq:MIrr}$$

The Gaussian curvature, in terms of $\L$, takes the form $K= e^{-2\L}( 1 -
\bl0\L)$ which, making use of Stokes' theorem, immediately gives the
conservation of the Euler number for $\Sur$:
      $$\chi_{\S} = {1\over 2\pi}\int_{S^2}{Ke^{2\L}\vol0} =2.$$

Singleton\R(Sin90b) gave an expression for the Bondi-Sachs mass of the
spacetime:
      \begin{eqnarray}%%
             M_B  ={m\over4\pi}\int_{S^2}{e^{3\L} \vol0} \label{eq:BMdef}
      \end{eqnarray}%%
which is manifestly positive and in fact is bounded below by
the irreducible mass, since by the H\"older inequality
      $$ \left(\int_{S^2}\vol0\/\right)^{1\over 3\/}
      \left(\int_{S^2} e^{3\Lambda}\vol0\/\right)^{2\over 3\/}
      \geq \int_{S^2} e^{2\Lambda}\vol0        \label{eq:massbound} $$

That the Bondi mass is monotonically decreasing can be shown
by differentiating \E(BMdef) and using the \RT equation:
\be
      \ddu M_B = -\frac{1}{4\pi}\int_{S^2}
       e^\L (\P0^2(e^{-\L}),_\zb),_\zb (\P0^2(e^{-\L}),_\z),_\z\vol0
%	e^\L \e0\e0(e^{-\L})\e0'\e0'(e^{-\L}) \vol0
	\quad \leq 0 \label{eq:BMdecrease}
\ee
These characteristics enabled Singleton\R(Sin90b) to show that the Bondi Mass
is a Lyapunov functional for the \RT evolution,  complementing the earlier
work of Luk\'acs et. al\R(LPPS) who  showed that the integral $\int_{S^2} K^2
e^{2\L}\vol0$ is a Lyapunov functional for  the \RT evolution.  These Lyapunov
functionals played a key role in Chru\'sciel's semi-global existence
proof.\R(Chr91)

\vfill\pagebreak
\section {The Past Apparent Horizon}
\subsection{General Remarks}
Apparent horizons in black hole spacetimes have been the subject of much study
in recent times.  The motivation for the study of  the future or past apparent
horizons is that they provide a locally  characterisable indication of the
presence of an event horizon (future)  or particle horizon (past).  Indeed, in
cases where the event horizon or  particle horizon cannot be properly defined,
the apparent horizon may be  the only useful definition of the ``surface'' of
the black hole or white hole.\R(Hay94)  This arguably is the case in the \RT
spacetimes, where $\Scri-$ cannot be  properly defined due to the instability
of the \RT equation in the negative $u$ direction.\R(Chr91)  Apparent horizons
have been widely used in numerically generated spacetimes as an indicator of
the presence of a black hole,\R(NKO84,CY90) in lieu of being able to detect
the event horizon in a local way\footnote{ In recent work, Anninos
\etal\R(Ann94) have been able to locate the event horizon in  numerically
generated spacetimes by integrating backwards from the stationary final state
of the black hole}.   It is believed that the presence of an apparent horizon
indicates the presence of a nearby event horizon (where it exists), and it has
been shown that for all stationary black  hole spacetimes the apparent and
event horizons coincide.\R(HE)  Recent work has shown that the apparent
horizon of a black hole has ``thermodynamic'' properties: in particular it has
been shown that a future apparent horizon can only increase in area, while a
past apparent horizon can only decrease in area, obeying a law like that of
the first law of thermodynamics.\R(Col92,Hay94)

The Robinson-Trautman spacetimes, as they settle down to the Schwarzschild
solution as $u\rightarrow\infty$, would be expected to have a well behaved
future apparent horizon coinciding with the event horizon.   Due to the
presence of outgoing radiation, we would expect the past  apparent horizon to
be more complex.   As a general (and non rigorous) consideration, it can be
seen quite clearly from the line element \E(metric) that the curvature
singularity at $r=0$ is a spacelike singularity i.e. that if the spacetime
were complete in the past null direction (if $\Scri-$ were complete) then
there would exist a particle horizon -- the past counterpart of the event
horizon (see \Fig(pd)).  Effectively, the past singularity being spacelike
indicates that no null geodesics from $\Scri-$ can reach the singularity, if
the spacetime is to be causally well-behaved.  Because the particle horizons
is a globally defined object, it does not exist in \RT spacetimes.  However,
the spacelike nature of the past singularity suggests that there should always
be a past apparent horizon in these spacetimes,  which we believe to be
causally well-behaved.

\begin{figure}[th]
\centering
\setlength{\unitlength}{0.007500in}%
\begin{picture}(0,0)%
\includegraphics{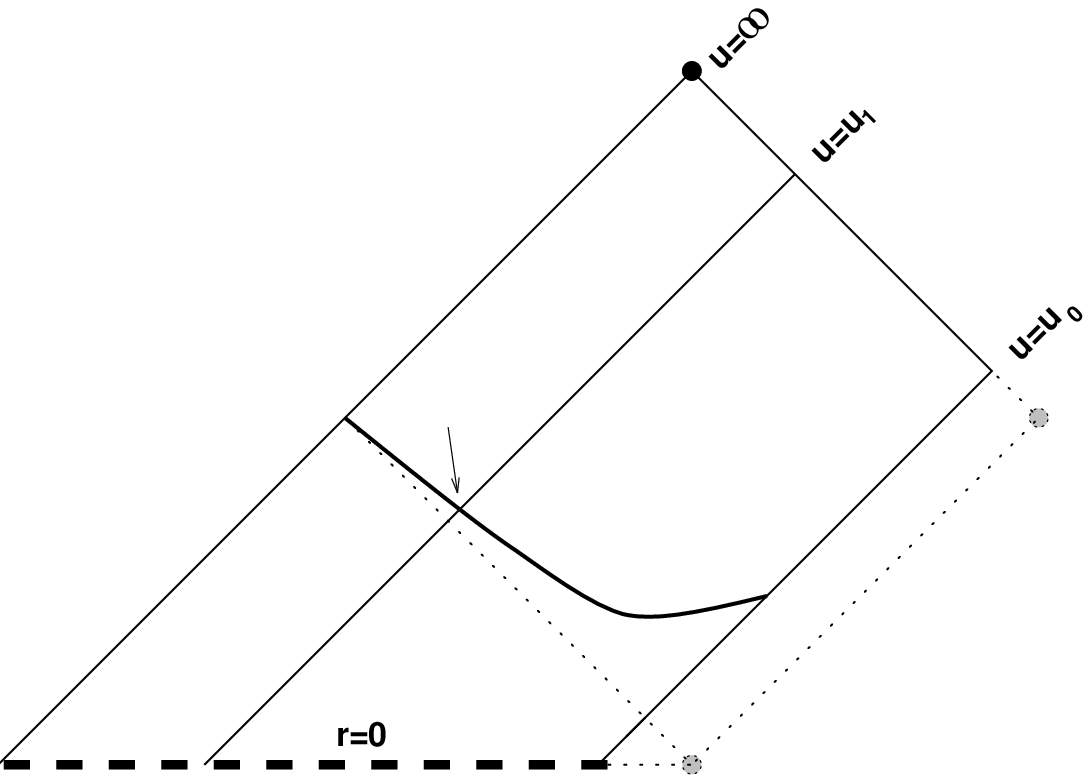}%
\end{picture}%
\begingroup\makeatletter\ifx\SetFigFont\undefined
% extract first six characters in \fmtname
\def\x#1#2#3#4#5#6#7\relax{\def\x{#1#2#3#4#5#6}}%
\expandafter\x\fmtname xxxxxx\relax \def\y{splain}%
\ifx\x\y   % LaTeX or SliTeX?
\gdef\SetFigFont#1#2#3{%
  \ifnum #1<17\tiny\else \ifnum #1<20\small\else
  \ifnum #1<24\normalsize\else \ifnum #1<29\large\else
  \ifnum #1<34\Large\else \ifnum #1<41\LARGE\else
     \huge\fi\fi\fi\fi\fi\fi
  \csname #3\endcsname}%
\else
\gdef\SetFigFont#1#2#3{\begingroup
  \count@#1\relax \ifnum 25<\count@\count@25\fi
  \def\x{\endgroup\@setsize\SetFigFont{#2pt}}%
  \expandafter\x
    \csname \romannumeral\the\count@ pt\expandafter\endcsname
    \csname @\romannumeral\the\count@ pt\endcsname
  \csname #3\endcsname}%
\fi
\fi\endgroup
\begin{picture}(570,409)(160,305)
\put(400,615){\makebox(0,0)[lb]{\smash{\SetFigFont{12}{14.4}{rm}$\H^{+}$}}}
\put(625,605){\makebox(0,0)[lb]{\smash{\SetFigFont{12}{14.4}{rm}$\Scri+$}}}
\put(530,690){\makebox(0,0)[lb]{\smash{\SetFigFont{12}{14.4}{rm}$i^{+}$}}}
\put(450,425){\makebox(0,0)[lb]{\smash{\SetFigFont{12}{14.4}{rm}$\H^{-}$}}}
\put(390,490){\makebox(0,0)[lb]{\smash{\SetFigFont{12}{14.4}{rm}
	${{\cal T}_{u_1}}^-$}}}
\put(620,365){\makebox(0,0)[lb]{\smash{\SetFigFont{12}{14.4}{rm}$(\Scri-$)}}}

\end{picture}
\caption{\ninerm Penrose diagram for the \RT spacetime. \label{f:pd}}
\end{figure}

\subsection{Apparent Horizon Equations}
The concept of an apparent horizon rests on the definition of marginally
trapped surface.  We define an {\em outer past marginally trapped surface},
${\Tm}$, to be a spacelike two surface on which the ingoing future directed
congruence of orthogonal  null geodesics has vanishing divergence, while the
outgoing future directed congruence of orthogonal null geodesics is
diverging.  We define the {\em outer past apparent horizon}, $\Hm$, to be a
hypersurface $r=\V(u,\z,\zb)$ such that its intersection with each $u=u_0$
slice is an outer marginally  past trapped two-surface $\Tmu$, i.e. $\Hm$ is
foliated by the surfaces $\Tmu$.   The past apparent horizon is illustrated in
\Fig(pd).

The equations describing the apparent horizon are constructed as follow.
Consider a null tetrad $(l^a, n^a, m^a, \b m^a)$ for
the spacetime with $l^a$ tangent to a congruence of outgoing
shearfree diverging null geodesics, and $m^a$ and $\b m^a$ lying
on a family of two surfaces  $\Tu$:
\begin{eqnarray}
       l^a &=&\partial_r  \nonumber\\
      n^a &=& \partial_u + \left[{P^2 \V_\z \V_\zb \over
      r^2} - H \right]\partial_r +
	   {P^2 \V_\z \over r^2}\partial_\z + {P^2 \V_\zb \over r^2}\partial_\zb
	   \nonumber\\ m^a &=&  {P\over r}\partial_\zb + {P\V_\zb
      \over r}\partial_r \nonumber \\
      \mb^a &=&  {P\over r}\partial_\z + {P\V_\z
      \over r}\partial_r \label{eq:tetrad}
 \end{eqnarray}

where $H$ is as in  \E(metric) and $r=\V(u,\z,\zb)$ defines the
surfaces $\Tu$.
The nonvanishing Newman-Penrose quantities are then given
by:

\begin{eqnarray}
      \rho &=& -{1\over r}\/,\;\;\;
      \tau = {\overline \pi} = {\overline \alpha}+\beta
      = -{P\Re,_{\bar \zeta} \over r^2}\/,\;\;\;
      \beta = -{P,_{\bar \zeta} \over 2r}\/,   \nonumber     \\
      \lambda &=& (\partial_\zeta +  \Re,_\zeta \partial_r)
      \left({P^2 \Re,_\zeta \over r^2}\right)\/,   \nonumber   \\
      \mu &=&  -{1 \over 2r} \left[ K
       - {2 \over r}(m + P^2 \Re,_{\zeta {\bar \zeta}})
      +{1 \over r^2}({2P^2 \Re,_\zeta \Re,_{\bar \zeta}})\right]\/,
      \nonumber       \\
      \gamma &=&  -{1 \over 2} \left[ (\ln P),_{u},-
      {1 \over r^2}(m + {P P,_\zeta \Re,_{\bar \zeta}}
                         - {P P,_{\bar \zeta} \Re,_\zeta})
      + {1 \over r^3}({2P^2 \Re,_\zeta \Re,_{\bar \zeta}}) \right]\/,
      \nonumber       \\
      \nu &=& {P \over r}(\partial_\zeta +  \Re,_\zeta \partial_r)
      \left(\Re,_u -r(\ln P),_u + {K \over 2} -{m\over r}
      +{P^2 \Re,_{\zeta} \Re,_{\bar \zeta} \over r^2}\right)\/,  \nonumber
\\
      \Psi_2 &=& -{m \over r^3}\/,    \nonumber    \\
      \Psi_3 &=& -{P K,_\zeta \over 2r^2}
                -{3 P m \Re,_\zeta \over r^4}
               \/,   \nonumber   \\
      \Psi_4 &=& -{[P^2 (\ln P),_{u \zeta}],_{\zeta} \over r}
               +{(P^2 K,_\zeta),_\zeta \over 2r^2}
               -{2P^2 K,_\zeta \Re,_\zeta \over r^3}
              -{6 P^2 m (\Re,_\zeta)^2 \over r^5}
        \label{eq:spcf}
\end{eqnarray}

{}From \E(tetrad) and \E(spcf) it can be seen that the null vector $l^a$ is
future directed, outward pointing, geodetic ($\kappa=0$), diverging ($\rho=-{1
\over r}$), null and orthogonal to the two-surface ${\cal T}_u$.  That is,
$l^a$ is tangent to the outgoing congruence of null geodesics normal to $\Tu$.
The tetrad vector $n^a$ will be tangent at $\Tu$ to the ingoing congruence of
null geodesics normal to $\Tu$.  The divergence of  $n^a$ is given by
$Re(\mu)$ ($= \mu$ in this case). Thus the divergence of the ingoing
congruence of null geodesics is given on $\Tu$ by $\w\mu$, where the tilde is
used here and subsequently to denote the restriction of an $r$-dependent
quantity to the hypersurface $r=\V(u,\z,\zb)$. If $\w\mu=0$, $\Tu$ are
marginally past-trapped two-surfaces $\Tmu$ and  the hypersurface
$r=\V(u,\z,\zb)$ is the past apparent horizon $\Hm$. This condition gives us
the first horizon equation:
      \begin{eqnarray}%%
            K - {2m \over \Re}  - \Delta(\ln \Re) = 0
             \label{eq:mu}
      \end{eqnarray}%%
This equation is attributed to Penrose\R(Pen73).  Tod\R(Tod89)
examined it in more detail, proving that \E(mu) has a unique
$C^{\infty}$ solution given a $C^{\infty}$ background $\Sur$,
and also that the surface $\Tmu$ defined by the solution is
in fact the outer boundary of past-trapped surfaces on $u=u_0$.

The other equations are found by examining the
embedding of the past apparent horizon in the spacetime; in particular
by considering the normal vector to the apparent horizon.

Let ${\N}_{a}$
be a one-form on ${\Hm}$ defined by
${\N}_{a} = -\Re,_u du + dr - \Re,_{\zeta} d\zeta - \Re,_{\bar\zeta}
d{\bar\zeta}\/$. Hence ${\N}^{a}:=g^{ab}{\N}_{b}$ is a
vector orthogonal to ${\Hm}$ and is given by
\begin{eqnarray}
{\N}^a &=& {\w n}^a  - \left(\V,_{u} + {\w{H}} +
{P^2 \Re,_{\zeta} \Re,_{\bar \zeta} \over \Re^2}\right){\w l}^a
                 \nonumber        \\
              &=& \partial_u - (\V,_{u} + 2{\w{H}})\partial_r +
{P^2 \Re,_{\bar \zeta} \over \Re^2}\partial_\zeta +
{P^2 \Re,_\zeta \over \Re^2}\partial_{\bar \zeta}
                 \label{eq:normal}
\end{eqnarray}

Since the complex null vectors
${\w m}^a$ and ${\w{\overline m}}^a$ are tangent to
the two-sufaces $\Tmu$ which foliate the hypersurface ${\Hm}$,
the null vectors ${\w l}^a$ and
${\w n}^a$ are orthogonal to $\Tmu$. It can be seen that
the vector
\begin{eqnarray}
      {\Z}^a &=& {\w n}^a  + \left(\V,_{u} + {\w{H}} +
      {P^2 \Re,_{\zeta} \Re,_{\bar \zeta} \over \Re^2}\right){\w l}^a
                       \nonumber                           \\
	  &=&\partial_u + \left(\V,_{u}+{2P^2 \Re,_\zeta \Re,_{\bar \zeta} \over
	  \Re^2} \right)\partial_r + {P^2 \Re,_{\bar \zeta} \over
	  \Re^2}\partial_\zeta + {P^2 \Re,_\zeta \over \Re^2}\partial_{\bar \zeta}
      \label{eq:triad}
\end{eqnarray}
is orthogonal to ${\N}^a$ and therefore is tangent
to the hypersurface ${\Hm}$.  The ``magnitude'' of
${\N}^a$ and ${\Z}^a$,
from (\ref{eq:normal}) and (\ref{eq:triad}), is given by
\begin{eqnarray}
      {\N}_a{\N}^a=-{\Z}_a{\Z}^a
      &=&-2\left(\V,_{u} + {\w{H}} +
      {P^2 \Re,_{\zeta} \Re,_{\bar \zeta} \over {\Re}^2}\right)
%     \nonumber\\
%      &=& -2\V\partial_u\left(\ln{\V\over P}\right) - K + {2m\over \V} -
%     {2P^2\V,_\z \V,_\zb\over R^2}
	\label{eq:length}
\end{eqnarray}

Since from \E(mu) ${\w{\mu}}=0$ on each $\Tu$,
the directional derivative of ${\w{\mu}}$ along the
vector ${\Z}^a$ tangent to ${\Hm}$ must vanish.
(\ref{eq:triad}) and (\ref{eq:length}) then imply
\begin{eqnarray}%%
      {\Z}^a \nabla_a \w{\mu} =
      {\w n}^a \nabla_a \w{\mu} -
      {1 \over 2}({\N}_b{\N}^b\,)
      {\w l}^a \nabla_a \w{\mu} = 0      \label{eq:dd}
\end{eqnarray}%%
\noindent Substituting \E(spcf) into
the Newman-Penrose equations\R(PR)
% $(4.11.12.a')$ and $(4.11.12f')$ in Penrose and Rindler\R(PR)$
gives

\begin{eqnarray}
{\w n}^a \nabla_a \w{\mu}&=&
{\w m}^a \nabla_a \w{\nu} +
\w{\nu}(-\w{\tau} + \w{\overline \alpha} +
3\w{\beta}) + \w{\pi} \w{\overline \nu} -
\w{\lambda} \w{\overline \lambda} -
2\w{\phi}_2 \w{\overline \phi}_2    \nonumber      \\
&=& -\partial_{\zeta}\!\left({P^2 \Re,_\zeta \over \Re^2}\right)
\partial_{\bar \zeta}\!\left({P^2 \Re,_{\bar\zeta} \over \Re^2}\right)
\nonumber\\
%-{P^2\over 2Q_{0}^2 \Re^2}\partial_{\zeta}\!\left(M-{2Q_{0}^2
%\over \Re}\right) %\partial_{\bar \zeta}\!\left(M-{2Q_{0}^2 \over \Re}\right)
\nonumber   \\
& &\qquad\qquad -{P^2 \over 2\Re}\left[\,\partial_{\zeta {\bar \zeta}}\!\left(
{{\N}_a{\N}^a \over \Re}\right) - ({\N}_a{\N}^a) \,\partial_{\zeta {\bar
\zeta}}\!\left({1 \over \Re}\right) \right]
\label{eq:np1}        \\
{\w l}^a \nabla_a \w{\mu}&=&
{\w m}^a \nabla_a \w{\pi} +
\w{\pi}(\w{\overline \pi} - \w{\overline \alpha} +
\w{\beta}) + \w{\Psi}_2    \nonumber     \\
&=&{P^2 \over \Re} \partial_{\zeta{\bar \zeta}}\left({1 \over \Re}\right)
- {m\over \Re^3}            \label{eq:np2}
\end{eqnarray}

In (\ref{eq:np1}) we have used
%$\w\nu= -\frac{1}{2\V}\edth'(N^a N_a)$ and $\w\lambda =
$\w{\nu}=-{P \over 2\Re} ({\N}_a{\N}^a),_\zeta\,$ and $\w\lambda =
(\frac{P^2\V,_\z}{\V^2}),_\z$.
Combine (\ref{eq:dd}), (\ref{eq:np1}) and (\ref{eq:np2})
to give

$$
      {1\over \Re}\Delta\left({{\N}_a{\N}^a \over \Re}\right)
      +2 \w{\Psi}_2({\N}_a{\N}^a)
      =  -4\w{\lambda} \w{\bar\lambda}  \label{eq:h3}
$$

To summarise, then, the apparent horizon in \af\RT
spacetimes can be described
by the following equations:
\begin{eqnarray}
      & &K - {2m\over \V} =\Delta(\ln\V) \label{eq:hor1}\\
	  & &{1\over \V}\Delta\left({\N_a \N^a\over \V}\right) - 2{m\over
	  \V^3}(\N_a \N^a) = - 4\w\lambda\w{\b\lambda} \label{eq:hor2}\\
	  & &\N_a
	  \N^a = - 2\V\left(\ln{\V\over P}\right),_u - K + {2m\over \V} +
      - {2P^2\V,_\z \V,_\zb \over \V^2}
       \label{eq:hor3}
\end{eqnarray}

The $Re(\t\mu)=0$ equation \E(hor1) is the primary equation required to locate
the horizon.  Tod's proof of existence and uniqueness of solutions for this
equation indicates that there  must be a marginally trapped two-surface on
each  slice $u=u_0$.  However, in order for $\Hm$, the union of these
surfaces, to  be considered an apparent horizon, we require that $\Hm$ be a
non-timelike hypersurface,  i.e. that $\N^a \N_a$ is always non-negative.  As
shown by Collins\R(Col92) and Hayward,\R(Hay94) this is equivalent to the
assumption that the directional derivative of the divergence of the ingoing
null geodesics in the direction of future pointing outgoing null geodesics is
negative.  In fact, Hayward makes this assumption a part of the definition of
an outer apparent horizon, and thus does not require a spacetime to be
asymptotically flat in order for the outer apparent horizon to be selected.
In the \RT spacetime, which is asymptotically flat, we can prove that this
assumption is not necessary; using the maximum principle and \E(hor2) it is
straightforward to show that $N^aN_a\geq0$ (details of the proof will be given
elsewhere).

\subsection{Properties of the horizon}
Equation \E(hor1) can be written in the form
     \begin{eqnarray}%%
         2me^{3\L} = e^\Phi (1-\bl0\Phi) \label{eq:hor4}
      \end{eqnarray}%%
where $e^\Phi=\V{e^\L}$. This gives a useful expression for the Gaussian
curvature of the apparent
horizon, $K_T = e^{-2\Phi}(1-\bl0\Phi)$.  It is then straightforward
to show that the marginally trapped surfaces have the topology of $S^2$:
    \begin{eqnarray}%%
        \chi_T = {1\over 2\pi}\int_{S^2} K_T e^{2\Phi}\vol0 =
        {1\over 2\pi}\int_{S^2} [1-\bl0\Phi]\vol0 = 2.
     \end{eqnarray}%%
Furthermore, because $K_T e^{3\Phi} = 2me^{3\L} $,
$K_T$ must be positive everywhere.  Also, this relationship gives rise
to some additional interesting integral quantities:

\begin{eqnarray}
	  \int_{S^2} me^{2\L} \vol0 = 4\pi m &\Leftrightarrow&\int_{S^2} K_T
	  e^{3\Phi-\L} \vol0 = 8\pi m\nonumber\\
	  \int_{S^2} me^{3\L-\Phi} \vol0 = 2\pi &\Leftrightarrow&\int_{S^2} K_T
	  e^{2\Phi} \vol0 = 4\pi \nonumber\\
	  \int_{S^2} me^{3\L} \vol0 \geq 0 \qquad&\Leftrightarrow&\int_{S^2} K_T
	  e^{3\Phi} \vol0 \geq 0 \nonumber\\
	  \ddu\int_{S^2} me^{3\L} \vol0 \leq 0 \qquad
	  &\Leftrightarrow&\ddu\int_{S^2} K_T e^{3\Phi} \vol0 \leq 0
      \label{eq:mKiden}
\end{eqnarray}

This is suggestive of the fact that the \RT equation can be rewritten entirely
 in terms of ``horizon quantities''.  From \E(hor4), we have

$$e^{2\L} = \left( {1\over 2m}e^\Phi(1-\bl0\Phi)\right)^{2\over3}$$
and
$$ \bl0\L = {1\over 3} \bl0(\Phi + ln(1-\bl0\Phi)).$$
Using the \RT equation in the form
\begin{eqnarray}%%
   (e^{2\L}),u = -{1\over 12m}\bl0(e^{-2\L}[1-\bl0\L]),  \nonumber
\end{eqnarray}%%
we can construct an equation in $\Phi$ alone.  While rather messy,
this formulation of the \RT equation is physically
interesting as it describes in some sense the
``evolution'' of the apparent horizon of the white hole. The term
``evolution'' is used loosely here, since we expect the
past apparent horizon to be non-timelike.

The properties of the area of the apparent horizon,
$\AT = \int_{S^2} e^{2\Phi} \vol0$, can also be
derived from the horizon equations, \E(hor1) to \E(hor3).
Tod proved the isoperimetric inequality, $ 16\pi M_B^2 \geq \AT$,
for the area of the horizon.\R(Tod86)
In fact, $ 16\pi M_B^2 \geq \AT \geq 16\pi m^2$, since, from the
H\"older Inequality and \E(mKiden),
\begin{eqnarray}
\left(\ints2 e^{2\Phi}\vol0\right)^{1\over3}
\left(\ints2 e^{3\L-\Phi}\vol0\right)^{2\over3} &\geq&\ints2
e^{2\L}\vol0\nonumber\\
\Rightarrow \qquad\qquad{\AT}^{1\over3} \left({2\pi\over m}\right)^{2\over3}
&\geq& 4\pi\nonumber\\
\Rightarrow \qquad\qquad\qquad\qquad \AT &\geq &16\pi m^2.\label{eq:arealb}
\end{eqnarray}

Since \E(hor2) implies that the apparent horizon must  be a non-timelike
hypersurface, i.e. that $N^aN_a\geq 0$ always, we can show quite easily that
the $\AT$ must always decrease. From \E(hor1) and \E(hor3), we have
\begin{eqnarray}%%
 (e^{2\Phi}),_u + \bl0(e^{\Phi-\L}) = - e^{\Phi+\L}(\N^a \N_a)
\end{eqnarray}%%
Thus
\begin{eqnarray}%%
   \ddu\AT = \ddu\ints2 e^{2\Phi}\vol0 = -\ints2 e^{\Phi+\L}(\N^a \N_a)\vol0
\leq 0.
\end{eqnarray}%%

Thus we have shown that the past apparent horizon $\Hm$ is a
non-timelike hypersurface foliated by past marginally trapped surfaces
$\Tmu$ whose surface area is a monotonically decreasing function of $u$.

\vfill\pagebreak
\section{Numerical modelling}
We are currently undertaking to solve the apparent horizon equations in
conjunction with the \RT equation  numerically, and to use the numerical model
to investigate further some of the properties of the apparent horizon.   Our
results at this stage demonstrate the nice properties of the horizon
equations.  We are initially solving the axisymmetric equations: thus the
system is 1+1 dimensional only.
A full discussion of the numerical results will be presented elsewhere,
but we give here a brief introduction to the work.

\begin{figure}[htbp]
\epsfxsize=0.95\hsize
\epsfbox{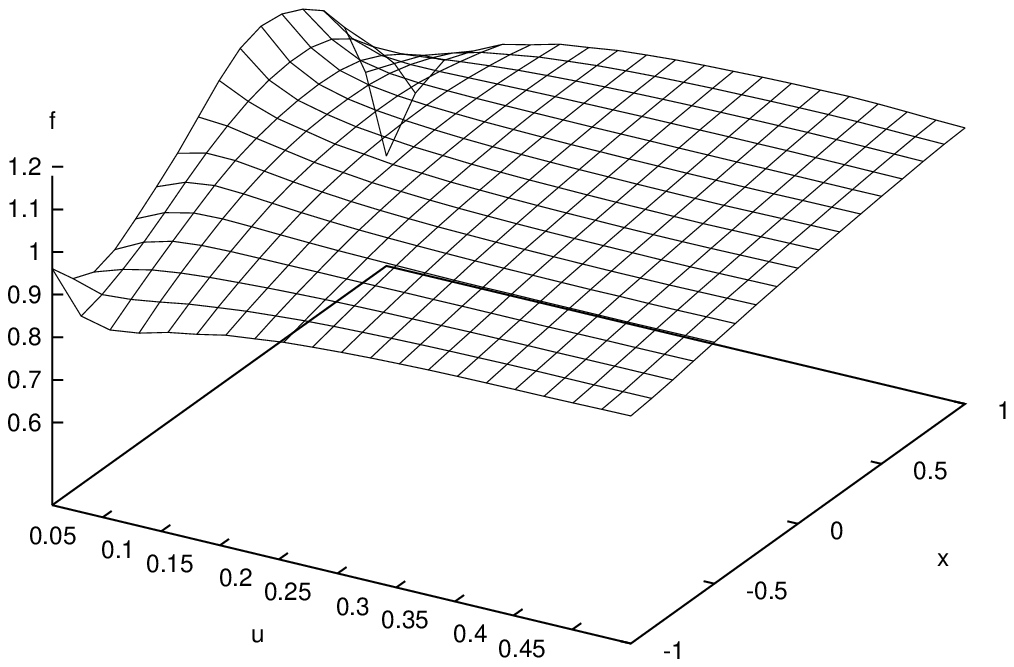}
\epsfbox{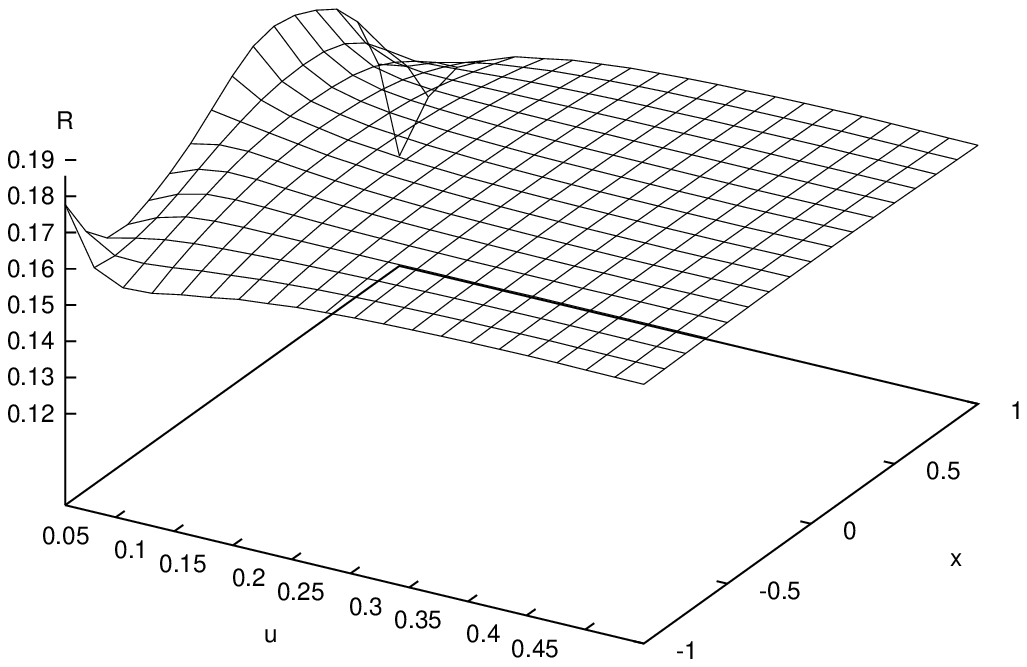}
\caption{\ninerm Numerical evolution of axisymmetric \RT spacetime and
apparent horizon. (a) $f(u,x) \equiv  {P\over P_0}$ (b)
$\V(u,x)$\label{f:poly}}
 \end{figure}

\begin{figure}[htb]
\centering
\begin{picture}(0,0)%
\includegraphics{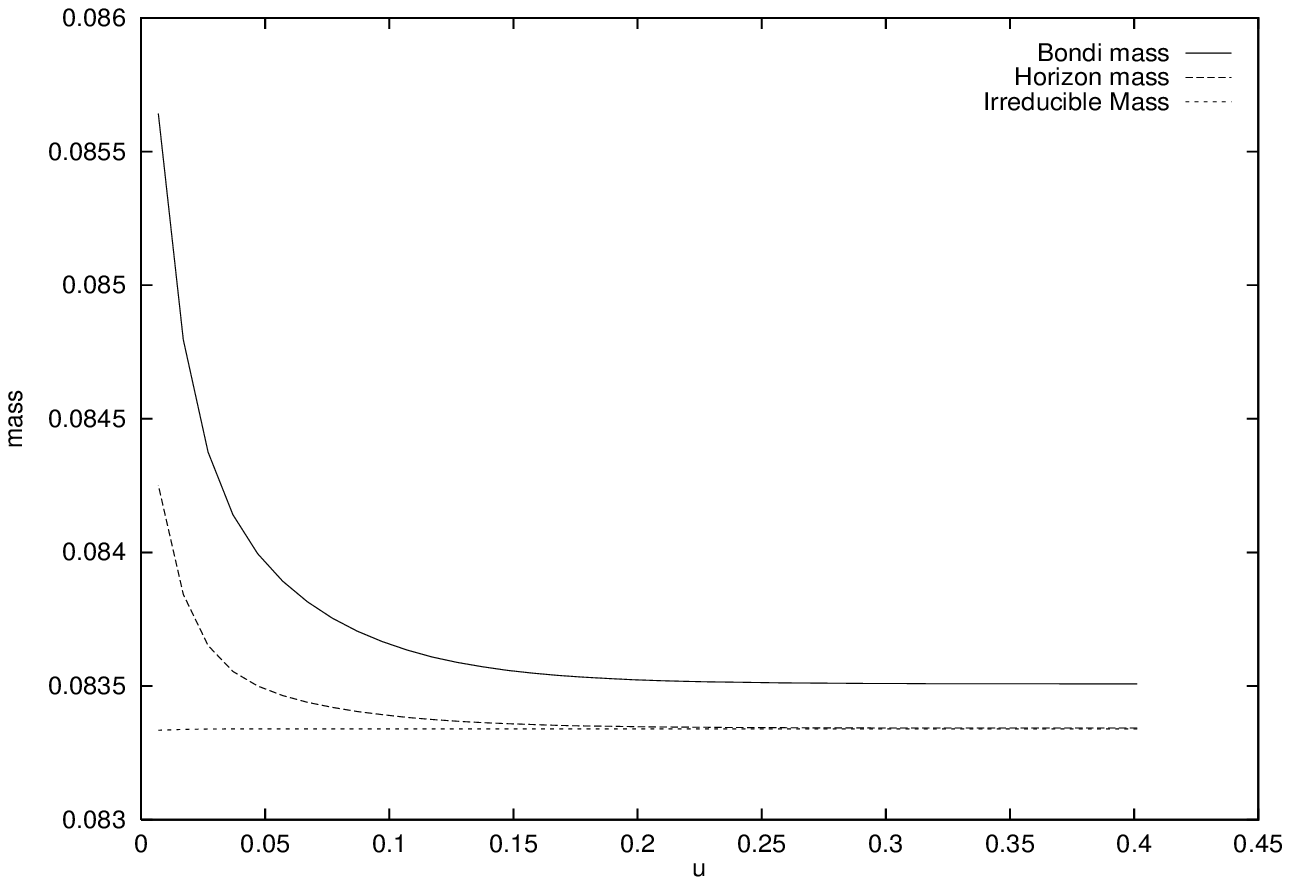}%
\end{picture}%
\setlength{\unitlength}{0.012500in}%
\begingroup\makeatletter\ifx\SetFigFont\undefined
% extract first six characters in \fmtname
\def\x#1#2#3#4#5#6#7\relax{\def\x{#1#2#3#4#5#6}}%
\expandafter\x\fmtname xxxxxx\relax \def\y{splain}%
\ifx\x\y   % LaTeX or SliTeX?
\gdef\SetFigFont#1#2#3{%
  \ifnum #1<17\tiny\else \ifnum #1<20\small\else
  \ifnum #1<24\normalsize\else \ifnum #1<29\large\else
  \ifnum #1<34\Large\else \ifnum #1<41\LARGE\else
     \huge\fi\fi\fi\fi\fi\fi
  \csname #3\endcsname}%
\else
\gdef\SetFigFont#1#2#3{\begingroup
  \count@#1\relax \ifnum 25<\count@\count@25\fi
  \def\x{\endgroup\@setsize\SetFigFont{#2pt}}%
  \expandafter\x
    \csname \romannumeral\the\count@ pt\expandafter\endcsname
    \csname @\romannumeral\the\count@ pt\endcsname
  \csname #3\endcsname}%
\fi
\fi\endgroup
\begin{picture}(420,280)(40,505)
\end{picture}
\vspace{0.5cm}
\caption{\ninerm Monotonic decay of mass quantities in the numerical evolution
of the \RT spacetime.\label{f:mass}}
\end{figure}

In order to find the apparent horizon, we first solve the \RT equation to find
the ``background spacetime''.  For this, we follow the work of
Singleton\R(Sin90b) and Prager and Lun,\R(Pra94) using exactly the
Crank-Nicolson algorithm described of the latter. Next, on each `time' slice
$u=u_0$, we solve \E(hor1) using Newton-Raphson iteration.  That is, for the
values of $P$ (or $\L$) determined by the \RT equation,  \E(hor1) is solved to
find the function $\V$ which describes the position of the marginally trapped
surface. Because the convergence of the Newton-Raphson method is entirely
dependent on making a good initial guess, we actually begin with the last
timestep of the evolution,  where the marginally trapped surface is known to
be close to $r=2m$, and solve for each time slice  in turn, going backwards to
$u=0$, taking as an initial guess for each slice the solution on the previous
slice.

The example shown has the initial condition  $f \equiv {P\over P_0} = 1. +
0.1Y_{1,0} - 0.2Y_{2,0} - 0.3Y_{3,0}$, where $Y_{\ell,0}$ are spherical
harmonics restricted to the axisymmetric case (i.e. Legendre polynomials).
For convenience we set $12m= 1$. \Fig(poly)(a) shows a plot of $f$ against
$x=\cos{\theta}$ evolving through  $u$.   It can be seen that the system
settles down to equilibrium ($P=P_0$)  very quickly.  Note that the
equilibrium solution includes a component of the first harmonic in this case,
so $f$ is of the form $f=a + bx$.  The function $\V$, representing the
position of the marginally trapped surface  $\Tmu$ is plotted in
\Fig(poly)(b), also showing very smooth behaviour. $\V$ is not subject to this
ambiguity in the final state, and settles down as  it should to $r=2m$, which
corresponds to the Schwarzschild equilibrium.

In \Fig(mass) the ``horizon mass'', defined as
$M_T = \sqrt{\frac{\AT}{16\pi}}$, is plotted as a function of $u$ with the
Bondi mass and the irreducible mass, demonstrating the monotonicity of the
 horizon area, and the inequalities described above (\E(arealb)).

\section{Concluding Remarks}
In this paper we have demonstrated the existence of the past apparent
horizon in the vacuum \af\RT spacetimes, and described some of its
properties.  Tod's proof of the existence of marginally past trapped
surfaces,\R(Tod89) together with our result that they foliate a
non-timelike hypersurface, allow us to call $\Hm$ a past apparent
horizon.
It is pleasing to find that the \RT spacetimes, which are the simplest
\af radiating spacetimes, exhibit a well behaved apparent horizon
structure.  In more general radiating spacetimes, we would not expect
the existence of apparent horizons to be guaranteed.  This is
demonstrated by examples such as the electrovac Robinson-Trautman\R(LC94)
or the vacuum Bondi-Sachs spacetimes.

We have also shown that the surface area of the past apparent horizon
decreases monotonically with the retarded time $u$.
This result is a particular example of more general theorems about
apparent horizon dynamics\R(Col92,Hay94).
However, we have not required all the assumptions used in the proof
of those theorems.

We have illustrated a method for describing apparent horizons analytically
based on the construction of a suitable null tetrad for the spacetime -  this
method could be easily applied to other cases.  In a future paper we will
discuss further the numerical solution of the horizon equations, as a means of
investigating the physics of these spacetimes.

\section{Acknowledgements}
We would like to thank Dan Prager for helpful discussions and
the Australian Research Council for financial support for this work.


\section{References}
\begin{thebibliography}{99.}
\bibitem{Ann94} Anninos P., Bernstein, D., Brandt, S., Libson, J., Masso, J.,
Seidel, E., Smarr, L., Suen, W.-M. and Walker, P. to appear in {\bibit General
Relativity {\bibit (MG7 Proceedings)}}, ed. R. Ruffini and M.Keiser \WS.95.
\bibitem{Chr91} Chru\'sciel, P.T. \CMP.137.289.91.
\bibitem{Col92} Collins, W. \PRD.45.495.92.
\bibitem{CY90} Cook, G. and York, J.W. \PRD.41.1077.90.
\bibitem{FN67} Foster, J. and Newman, E.T. \JMP.18.189.67.
\bibitem{GHHP} Gibbons, G., Hawking, S., Horowitz, G., and Perry, M.
\CMP.88.295.83.
\bibitem{Hay94} Hayward, S. A. \PRD.49.6467.94.
\bibitem{HE} Hawking, S.W. and Ellis, G.F.R. {\bibit The Large Scale Structure
of Space-time\/} \CUP.73.
\bibitem{HoePer} Hoenselaers, C. and Perj\'es, Z. \CQG.10.375.93.
\bibitem{LC94} Lun, A.W.-C. and Chow, E.W.M., to appear in {\bibit Int. J.
Mod. Phys.}
\bibitem{LPPS} Luk\'acs, B., Perj\'es, Z., Porter, J. and Sebesty\'en, \'A.
\GRG.16.691.84.
%\bibitem{Mal94} Malec, E. \PRD.49.6475.94.
\bibitem{NKO84} Nakamura, T.,  Kojima, Y. and Oohara, K. \PLet.106A.235.84.
\bibitem{Pen73} Penrose, R. \ANYAS.224.115.73.
\bibitem{Per89} Perj\'es, Z. in {\bibit Proceedings of the 5th Marcel Grossman
Meeting}, ed. D.G.Blair and M.J.Buckingham \WS.89.
\bibitem{PR} Penrose, R. and Rindler, W., {\bibit Spinors and Space-time},
Vol. 1 \CUP.84.
\bibitem{Pra94} Prager, D. and Lun, A.W.-C., to appear in {\bibit General
Relativity {\bibit (MG7 Proceedings)}}, ed. R. Ruffini and M.Keiser \WS.95.
\bibitem{RT60} Robinson, I., and Trautman, A. \PRL.4.431.60.
\bibitem{Ren88} Rendall, A.D. \CQG.5.1339.88.
\bibitem{Sch88} Schmidt, B. \GRG.20.65.88.
\bibitem{Sin90a} Singleton, D. \CQG.7.1333.90.
\bibitem{Sin90b} Singleton, D. (1990) Ph.D. Thesis, Monash University.  Also
in  {\bibit Proceedings of the 5th Marcel Grossman Meeting}, ed. D.G.Blair and
M.J.Buckingham \WS.89.
\bibitem{Tod86} Tod, K.P. \CQG.3.1169.86.
\bibitem{Tod89} Tod, K.P. \CQG.6.1159.89.
\bibitem{Van87} Vandyck, M.A.J. \CQG.4.759.87.
\end{thebibliography}
\end{document}